\documentclass[aps, twocolumn, showpacs, nofootinbib, PRB]{revtex4-1}
\usepackage{float}
\usepackage{amsmath}
\usepackage{graphicx}
\usepackage{hyperref}
\usepackage{color}
\usepackage{amssymb}
\usepackage{amsfonts}
\usepackage{lipsum,appendix}
\usepackage{amsthm}
\newcommand{\be}{\begin{equation}} 
\newcommand{\ee}{\end{equation}} 
\newcommand{\bea}{\begin{eqnarray}} 
\newcommand{\eea}{\end{eqnarray}} 
\newcommand{\bqa}{\begin{eqnarray}}
\newcommand{\eqa}{\end{eqnarray}}
\newcommand{\mb}{\mathbf}
\newcommand{\mc}{\mathcal}
\newcommand{\nn}{\nonumber \\}

\newcommand{\w}{\omega}

\newcommand{\lt}{\left}    
\newcommand{\rt}{\right}    
\newcommand{\figdrag}
{\begin{figure}[htbp]
        \centering
        \includegraphics[angle=0,width=8cm]{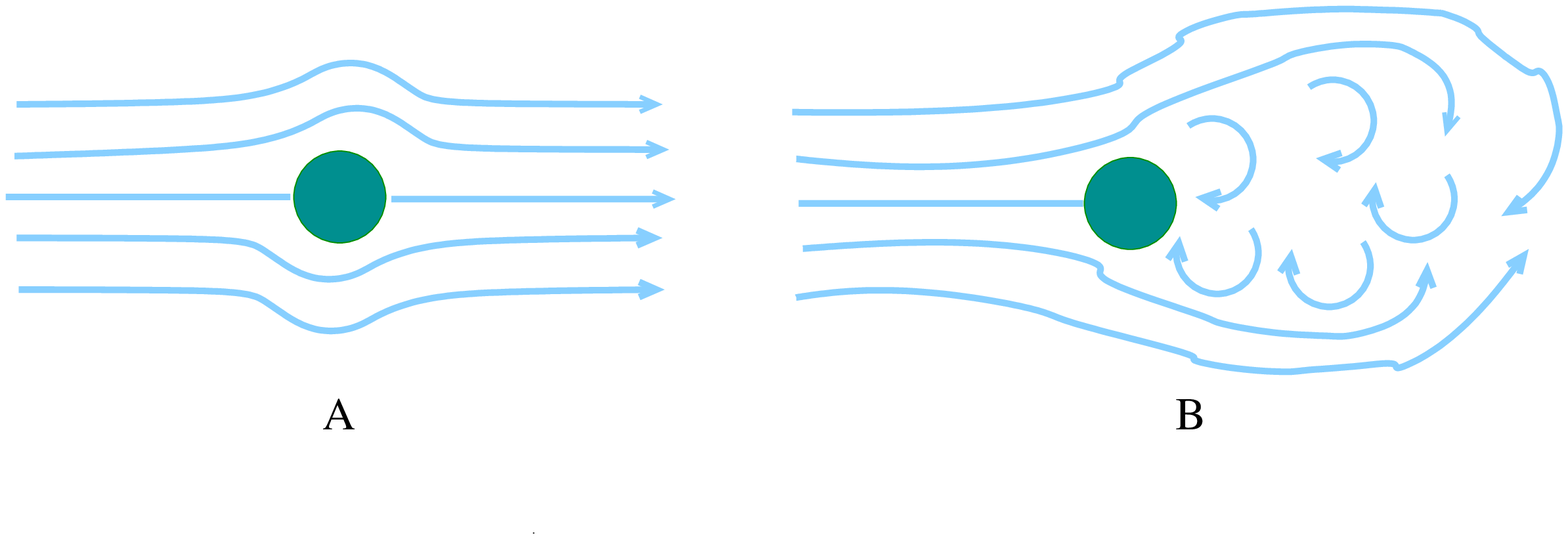}
           \caption{In Fig. 1A we see  viscous drag from the fluid molecules during a streamline flow, i.e. when an external particle of macroscopic size moves slowly inside the fluid. In this case the time scales corresponding to the macro-particle and the fluid particles are nicely separable. In Fig. 1B a turbulence sets in during a faster motion of a particle within fluid where such separation of scales is not possible.}
           \label{fig:figdrag}
	\end{figure}
}
\newcommand{\figproj}
{\begin{figure}[htbp]
        \centering
        \includegraphics[angle=0,width=6.5cm]{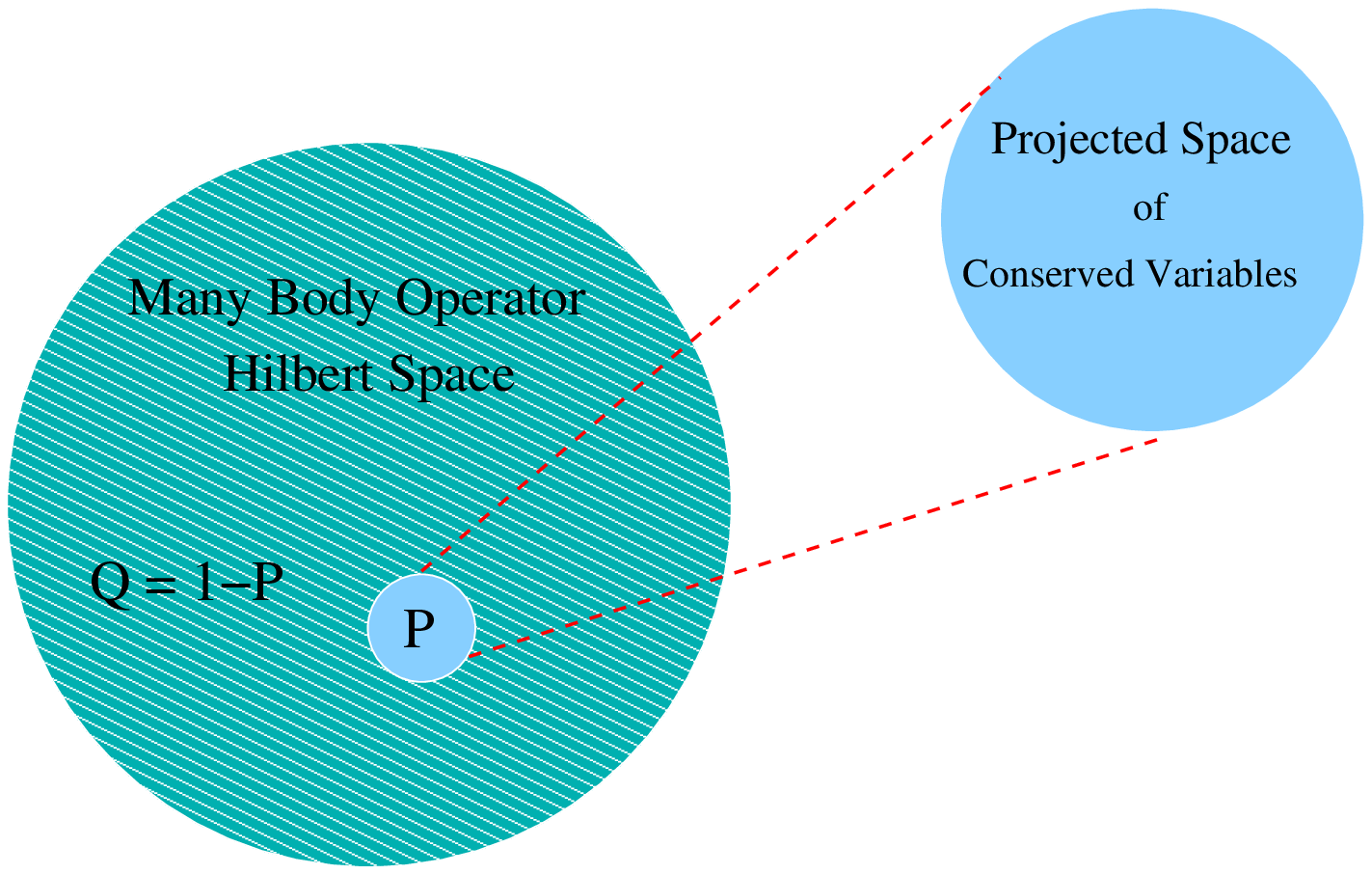}
           \caption{A schematic representation of the idea of projection in the memory function formalism. Here the full big circle is the total many body operator Hilbert Space and is equivalent to identity. The  Projection of the full many body states defined by 
           few operators representing conserved variables residing in the region $P$. On the other hand the incoherent degrees of freedom lives in the part of the Hilbert space defined by $\mathbb{I}-Q$.}
           \label{fig:figproj}
	\end{figure}
} 

\newcommand{\figT}
{\begin{figure}[htbp]
        \centering
        \includegraphics[angle=0,width=8cm]{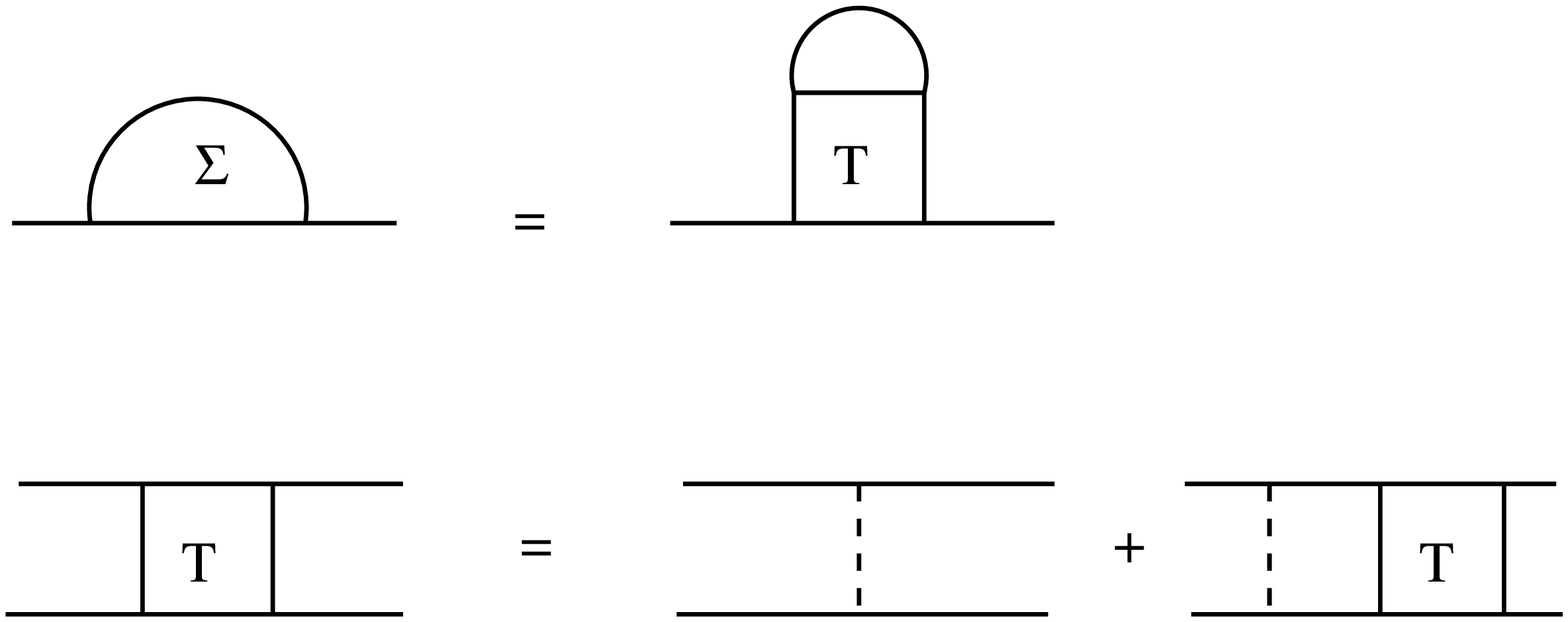}
           \caption{A diagrammatic description of the relation between the single particle self energy and the T-martix for electron-Boson interaction is shown here. The solid lines represents the electron propagator while the dashed lines represents Bosonic propagators.  }
           \label{fig:figT}
	\end{figure}
}\newcommand{\figbrown}
{\begin{figure}[htbp]
        \centering
        \includegraphics[angle=0,width=7cm]{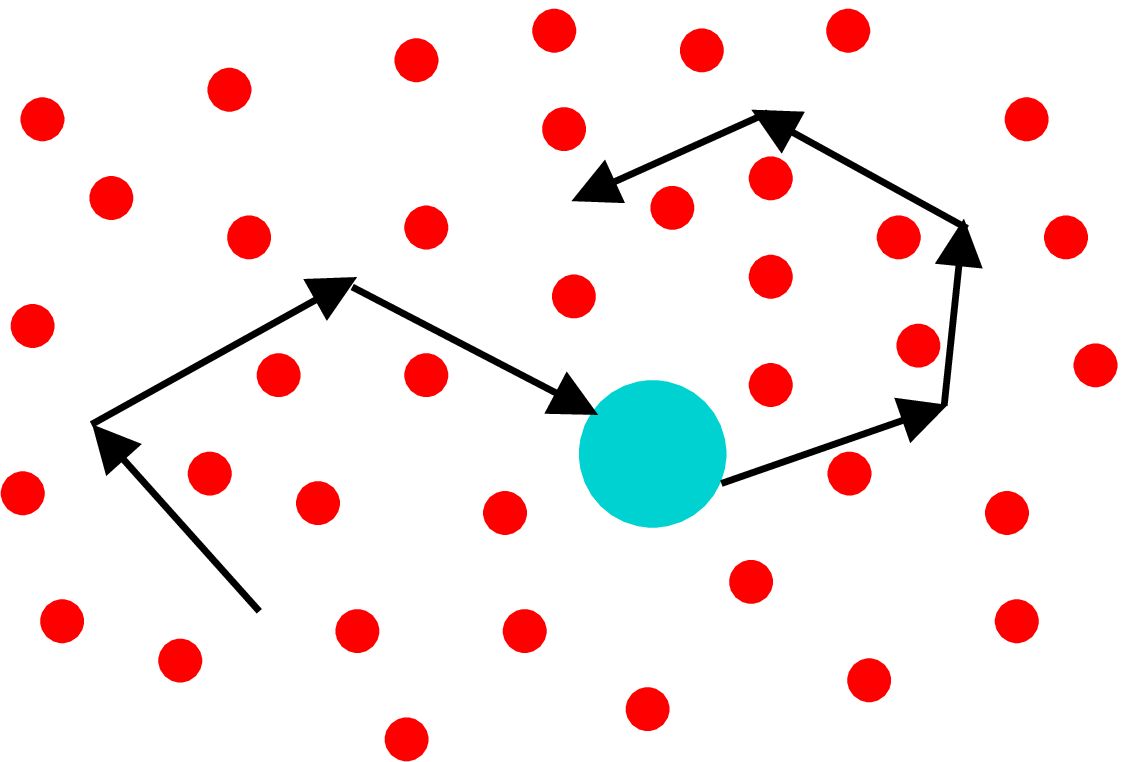}
           \caption{Motion of a Brownian particle (Big circle) moving in a fluid. Red dots represents small fluid molecules and they exhibit faster movements. The Brownian particle collides with the fluid particles and follows a zigzag path.}
           \label{fig:figbrown}
	\end{figure}
}
\usepackage{natbib}
\begin{document} 
\baselineskip 12pt 
\title {Memory Function Approach to Correlated Electron Transport:\\ A Comprehensive Review }
\author{Nabyendu Das$^1$, Pankaj Bhalla$^{1,2}$, Navinder Singh$^1$}
\affiliation{$^1$Theoretical Physics Division, Physical Research Laboratory, Ahmedabad-380009, India.\\
$^2$Indian Institute of Technology, Gandhinagar-382424, India.} 
\date{\today}
\begin{abstract} Memory function formalism or projection operator technique is an extremely useful method to study the transport and optical properties of various condensed matter systems. A recent revival of its uses in various correlated electronic systems is being observed. It is being used and discussed in various contexts, ranging from non-equilibrium dynamics to the optical properties of various strongly correlated systems such as high temperature superconductors. However, a detailed discussion on this method, starting from its origin to its present day applications at one place is lacking. In this article we attempt a comprehensive review of the memory function approach focusing on its uses in studying the dynamics and the transport properties of correlated electronic systems.
\end{abstract}
\maketitle 
\section{Motivation} Condensed matter  physics deals with the collective phenomena that emerge out of the mutual interactions between a large number of particles. Many of them are novel, i.e. are beyond the realm of pre-existent theories and almost none of them can have first principle microscopic explanations. Understanding such novel complex cooperative phenomena requires new physical ideas such as spontaneous symmetry breaking, Goldstone modes, renormalization of physical parameters etc.\cite{pwa} The basic principle of all such descriptions is adapting an effective description by separating the low energy slow degrees of freedom from the high energy faster degrees of freedom in a system of macroscopically large number of particles to study its low energy and long wavelength properties. Successes of such theories depend on how accurately one can develop an effective description of the effects of a large number of faster degrees of freedom on a few slower degrees of freedom which are experimentally probed in a certain physical system. As a result these theories are dependent on some common aspects of the various systems, e.g. energy and each of them works well for a certain class of physical systems\cite{pwa_book}.

Advent of the memory function formalism is a major theoretical progress in this line of thought. It was developed and is being used to study the dynamics and the transport properties of various complex many body systems\cite{zwanzig_61, zwanzig_61a, zwanzig_book, mori_65, mori_65-2, harp_70}. This technique relies on the idea of separating the slow and the faster degrees of freedom in a physical system and to systematically calculate the effects of the later on the former. Separation of scales  is a very familiar and essential concept in studying various physical 
systems. It is suitable in systems having finite number of slow modes related to the dynamics of conserved variables and or the broken symmetry variables. Their studies are often termed as hydrodynamics and those slow modes are called hydrodynamic modes and soft modes respectively\cite{forster_95}. 
\figdrag

To illustrate the idea of separating the scales, we can choose a simple example, a particle moving in a fluid. In this case, when the particle moves, the fluid particles oppose its motion as depicted in Fig.\ref{fig:figdrag}. An attempt to build up a microscopic theory for this motion will require a Hamiltonian that describes the Coulomb interactions between all the atoms and electrons present in the total system. Then one can try to solve Poisson or Schr\"odinger equation respectively depending on whether the system is classical or quantum in nature. Such an microscopic attempt is not only impossible, also it is too complicated to capture the essential physical description of the system\cite{pwa}. On the other hand we can build up a simple description  without compromising with the basic physics as follows. If the moving particle is macroscopically large and if the velocity of the center of mass  is small compared to the velocity of the fluid molecules, we can separate or project out 
the center of mass coordinates from the rest of the degrees of freedom of the total system. In such a situation, we can write its effective equation of motion where the  effects of the rest of the degrees of freedom can be termed as ``molecular drag'' on its motion. The later can be incorporated through a drag force. 
This leads to a very simple and well known equation of motion of the dragged particle of the form\cite{kubo_book},
\bea 
\mb{\ddot{R}}-\gamma \mb{\dot{R}}+\mb{F}=0.
\label{eq:drag}
\eea 
Here $\mb{R}$ is the position vector of the center of mass of the macroscopic particle of unit mass, $\mb{\dot{R}}$ and $\mb{\ddot{R}}$ represent its time derivative or the velocity and the acceleration respectively and $\mb{F}$ is the external force. This is indeed a major simplification of a very complex system. The parameter $\gamma$ is termed as friction coefficient, viscous coefficient etc. depending on the contexts. It describes dissipation or the flow of energy and or momentum from the coherent to the incoherent degrees of freedom in a system. It can also be space and time dependent. However when the velocity of the particle becomes comparable to that of the fluid molecules, as seen in the part B of the Fig.\ref{fig:figdrag}, turbulence sets in and the idea of separation of scales does not remain obvious. 

Many such examples can be found in the vast literature on complex systems both in the classical and quantum domains. It is to be noted that there is no concept of dissipation in the microscopic principles. In an effective description of a physical system, we observe the system within our desired or convenient time scale and length scale. We thus ignore the complete distribution of energy which occurs over a larger time and length scales. Hence we effectively observe dissipation of the momentum or energy of the particle  as a result of the interaction with other fast degrees of freedom. The same picture emerges in various effective descriptions of interacting systems and the study of the low energy properties becomes synonymous to finding out the generalized dissipative constant or the scattering rates of the collective excitations. We will see in the later sections that the   {\it Memory function approach} deals with systematic evaluations of the generalized dissipative constant($\gamma$ in Eqn. \ref{eq:drag}) related to the dynamics of generalized slow variables.

In this review, we will first address the general aspects of the Memory function formalism in section \ref{sec:formalism}. We elaborate it further in section \ref{sec:GLE} and in section \ref{sec:CF}. In those sections we present derivation of the generalized Langevin equation within this formalism and present a continued fraction representation of the dynamic correlation respectively. In section \ref{sec:ET} we discuss its applications in various electronic systems. Finally in section \ref{sec:diss}, we conclude.   
\section{Projectors and Memory Functions}
\label{sec:formalism}
 Memory function technique was introduced by  Zwanzig and Mori\cite{zwanzig_61, zwanzig_61a, zwanzig_book, mori_65, mori_65-2}. The mathematical formalism used for systematic calculation of the memory function is also called projection operator method. Following Ref.\cite{forster_95} and \cite{fulde_12} we summarize the  idea of using projectors and  the general mathematical setup for calculating the memory function. Let us start with a  many body system having macroscopically large number, i.e. an Avogrado number ($\sim 10^{23}/cc$) of degrees of freedom and look for its macroscopic properties. Our system can be both classical and quantum in nature. A classical system is described by a set of variables comprise of the particles position and momentum variables. Such a set of position and momentum variables is called phase space. In the systems obeying quantum mechanics a set of linearly independent operators do the same job. In that case a mathematical space containing the set of operators forms a Hilbert space as  depicted pictorially in the big green circle in Fig.\ref{fig:figproj}. 
\figproj
Now understanding the low energy consequences of the interactions between such a large number of variables or operators is extremely complicated, if not impossible. We need  methods which eliminate the technical difficulties but capture the correct low energy physics. 

Basic principle of the memory function formalism is as follows. Suppose we are interested in studying the center of mass motion of a
system of $N$ number of particles. Then we separate or project out the center of mass variable from the others. Here the center of mass variable is a macroscopic variable and is defined as a linear combination of the microscopic variables. Now in memory function formalism, it is shown that the effects of the rest of the microscopic variables on the dynamics of the macroscopic variable can be estimated systematically and is cast in a so called {\it Memory function} in different systems\cite{pires_04, khamzin_12, viswanath_book, schimacher_14, buishvili_86}. The reason behind the use of the term ``memory'' will be discussed in details in the next section. The above discussion is applicable to the quantum systems also, except the fact that the classical variables will be replaced by operators. Since we discuss this formalism in context of the electronic systems, we invoke quantum mechanics from the very beginning and work with operator language henceforth.

Let us consider an observable represented by an operator $A$ obeying the Hamiltonian dynamics. To determine its dynamics, we define the Liouville operator associated with the Hamiltonian. Mathematically the later is called a super-operator as it acts on operators and produces a new operator. For a system with a given Hamiltonian $H$, it is defined as,
\bea 
\mc{L}A=[H,A]=-i\frac{dA}{dt}.
\label{eq:liou}
\eea  
Here  $[\,\,,\,\,]$ is the commutator between two operators. From Eqn.\ref{eq:liou}, we see that an operator evolves with time as,
\be 
A(t)=e^{i\mc{L}t}A(0).
\ee 
 In a many body system we need to quantify the correlation between various physical quantities represented by various operators $A_i$. We can express their correlation in terms of a correlation function matrix $\mc{R}(z)$, defined in frequency space with matrix elements as,
\bea 
R_{ij}(z) &=& i\int dt e^{izt}\langle   A_i(t)\vert A_j \rangle =i\int dt e^{i(z-\mc{L})t}\langle   A_i  \vert A_j \rangle \nn
&=&\langle   A_i \vert \frac{1}{z-\mc{L}}A_j \rangle,\,\, z=\omega + i\eta .
\eea 
 Here $\eta\rightarrow 0^+$ is a small positive number, which assures causality. Evaluation of $R_{ij}(z)$ is a many body problem and is in general complicated. To simplify the evaluation of the above expression, we invoke the principle of memory function formalism and introduce a projection operator defined as follows\cite{forster_95, fulde_12},
\bea 
P&=&\sum_{ij}\vert A_i \rangle\chi_{ij}^{-1}\langle  A_j\vert ,\,\,\, \chi_{ij}=\langle  A_i \vert A_j \rangle ,\nn
&=&  \mathbb{I}-Q.
\eea
 Here $P$ separates the operator $A_i$s, corresponding to the observed macroscopic quantity, from the rest of the microscopic degrees of freedom and the act of $Q$ is just the opposite. 
A generic projection operator should have the following properties.
\be
P^2=P, \,\, PQ=QP=0, \,\, \rm{etc.}
\ee 
We also introduce a decomposition $\mc{L}=\mc{L}P+\mc{L}Q$ and introduce the identity
\be
\frac{1}{X+Y}=\frac{1}{X}-\frac{1}{X}Y\frac{1}{X+Y}.
\ee
Using the above identity the expression for the time dependent correlation function becomes
\be
R_{ij}(z)=\langle  A_i\vert \left\{\frac{1}{z-\mc{L}Q}+\frac{1}{z-\mc{L}Q}\mc{L}P\frac{1}{z-\mc{L}} \right\}\vert A_j \rangle.
\ee
Since $Q\vert A_j \rangle=0$, we can simplify the first term in the right hand side of the above expression as,
\be
\langle  A_i\vert \frac{1}{z-\mc{L}Q}\vert  A_j \rangle=\frac{1}{z}\langle  A_i \vert A_j \rangle=\frac{1}{z}\chi_{ij}.
\ee
 Therefore the expression for the correlator can be re-written as,
\be 
R_{ij}(z)=\frac{1}{z}\chi_{ij}+\sum_{lm}\langle  A_i \vert \frac{1}{z-\mc{L}Q } \mc{L} A_l\rangle \chi^{-1}_{lm}R_{mj}.
\ee
 Above expression can be cast in a matrix notation as follows, 
\bea 
\left(z\mathbb{I}-\mc{K}\chi^{-1}\right)\mc{R}=\chi.
\eea 
 The matrix $\mc{K}$ has the elements defined as,
\bea 
K_{il}=\langle  A_i \vert \frac{z}{z-\mc{L}Q} \mc{L} A_l \rangle.
\eea 
Elements of $\mc{K}$ can be decomposed as, 
\bea 
K_{il}=\langle  A_i \vert \mc{L} A_l \rangle +\langle  A_i \vert \mc{L}Q\frac{1}{z-\mc{L}Q} \mc{L} A_l \rangle.
\eea 
 The first part of the above expression is called the frequency matrix and is given as,
\bea 
\mc{L}_{il}=\langle  A_i \vert \mc{L} A_l \rangle.
\eea 
Remaining part contains the effects of the faster degrees of freedom residing in the un-projected part of the Hilbert space and is termed as the memory matrix. It is defined as,
\bea 
M_{il}=\langle  A_i \vert \mc{L}Q\frac{1}{z-Q\mc{L}Q} Q\mc{L} A_l \rangle.
\label{eq:memory}
\eea 
The relation $Q^2=Q$ is used to write it in a symmetric form. This form is very instructive. The above expression tells that the memory function is
defined in terms of the un-projected part of the $\vert \dot{A} \rangle=\mc{L} \vert A \rangle$ and the  un-projected part of the Liouville operator $\mc{L}$, i.e. $Q\mc{L}Q$. The projected degrees of freedom during their slow dynamics, can not keep track of the movements of the fast unprojected part and treat the later as incoherent excitations. Since the memory function consists of the unprojected degrees of freedom only, it  describes the effects of the incoherent excitations on the low energy excitations in a system and accounts for the dissipation in the slow degrees of freedom.
 Using the above expressions the correlator between different components of $A$ can be written in a compact notation as,
\bea 
\mc{R}(z)=\frac{1}{z\mathbb{I}-\left[ \mc{L}+M(z)\right]\chi^{-1}}\chi.
\label{eq:corrl}
\eea 
  Writting in terms of the matrix elements, it takes the form,
\bea 
\sum_l\left(z\delta_{il}-\sum_s \left[\mc{\mc{L}}_{is}+M_{is}\right]\chi^{-1}_{sl}\right)R_{lj}(z)=\chi_{ij}.
\label{eq:mem0}
\eea  
Here we see that any given correlation function can be written in terms of the corresponding memory matrix. This completes the general description of the memory function formalism. The use of it to study the electronic transport will be discussed in later sections.
\section{Generalized Langevin Equation}
\label{sec:GLE}
Before using this formalism in case of electronic transport, let us elaborate its physical contents in more detail in context of an well known physical system. Consider the dynamics of a system where few macroscopic slow degrees of freedom are immersed in and interacting with a soup of fast microscopic degrees of freedom. The famous Brownian motion is such an example\cite{brown_28, forster_95}. Here a particle is suspended in a fluid and it collides with the fluid molecules. As a result it follows a zigzag trajectory as shown in Fig.\ref{fig:figbrown}.
\figbrown\\
To explain such a phenomena, in the classical limit a scenario that the particle is experiencing ``some random force'',  is adopted.  As a result, for the particle a Newtonian equation of motion  with phenomenological random force, mimicking the kicks from the fluid particles can be written. Such an equation of motion is called Langevin equation\cite{langevin_08} and for the simplest case, in one dimension it takes the following form. 
\be 
\frac{d^2}{dt^2}  R (t)=\gamma\frac{d}{dt}  R (t)+f(t).
\ee
Here $f(t)$ is a random force of ``white noise'' type, i.e. with correlation $\langle   f(t)f(t')\rangle =\gamma\delta(t-t')$ and the mass of the particle is assumed to be unity. This correlation has a delta function in time structure and thus it is frequency independent. Such a random force description is used quite often and is highly successful in explaining various complex phenomena. How such a probabilistic picture emerges from the microscopic interactions which are deterministic in nature, can be addressed within the memory function formalism\cite{darve_09}.

Let us consider a time dependent operator $A(t)$ which mimic the velocity of a Brownian particle and consider its time evolution.
The corresponding Liouville equation is given as\cite{karasudani_79},
\be
\frac{d}{dt} A(t)=i\mc{L}A(t),\,\, A(t)=e^{i\mc{L}t}A(0).
\label{eq:GL1}
\ee
The time derivative of the operator $A(t)$ can be written in terms of its initial value $A(0)$ as,
\be 
\frac{d}{dt} A(t)=i\mc{L}e^{i\mc{L}t}A(0).
\ee
Now we introduce the projection operator acting on an operator $B$ as,
\bea
PB=\frac{\langle   A,B\rangle }{\lt\langle   A, A\rt\rangle }A,\,\ P^2=P.
\eea 
We insert the identity $ \mathbb{I}=P+Q$ in Eqn.\ref{eq:GL1}, and get,
\be 
\frac{d}{dt} A(t)=ie^{i\mc{L}t}(P+Q)\mc{L}A(0).
\label{eq:PQ}
\ee
Now onwards we drop the argument 0 from $A(0)$ for convenience and thus the first term in the above expression can be evaluated as,
\bea 
 e^{i\mc{L}t}P\mc{L}A =\frac{\langle   \mc{L}A,A\rangle }{\langle   A, A\rangle }e^{i\mc{L}t}A = \Omega e^{i\mc{L}t}A.
\eea 
Here $\Omega=\frac{\langle   \mc{L}A,A\rangle }{\langle   A, A\rangle }$ is the frequency matrix. To evaluate the second term we introduce the identity,
\bea 
e^{i\mc{L}t}&=&e^{i(P+Q)\mc{L}t}\nn
&=&e^{iQ\mc{L}t}+\int_0^t d\tau e^{i\mc{L}(t-\tau)}iP\mc{L}e^{iQ\mc{L}t}.
\eea 
Applying it to the second term of the Eqn. \ref{eq:PQ}, we get,
\bea 
e^{i\mc{L}t}iQ\mc{L}A&=&e^{iQ\mc{L}t}iQ\mc{L}A\nn
&&+\int_0^t d\tau e^{i\mc{L}(t-\tau)}iP\mc{L}e^{iQ\mc{L}t}iQ\mc{L}A.
\eea
We call the first term of the above equation a ``force'' which is given as,
\bea 
f(t)=e^{iQ\mc{L}t}iQ\mc{L}A=e^{iQ\mc{L}t}f(0).
\eea 
Here $f(0)$ is a ``force'' inserted on the slow variable by the incoherent degrees of freedom and $f(t)$ is formers time propagation. The other term can be evaluated as follows,
\bea 
I_2&=&\int_0^t d\tau e^{i\mc{L}(t-\tau)}iP\mc{L}e^{iQ\mc{L}t}iQ\mc{L}A\nn
&=&\int_0^t d\tau e^{i\mc{L}(t-\tau)}iP\mc{L}f(t)\nn
&=&\int_0^t d\tau e^{i\mc{L}(t-\tau)}i\frac{\langle   \mc{L}f(t),A\rangle }{\langle   A, A\rangle }A\nn
&=&\int_0^t d\tau e^{i\mc{L}(t-\tau)}i\frac{\langle   \mc{L}Qf(t),A\rangle }{\langle   A, A\rangle }A.
\eea 
Since both $Q$ and $\mc{L}$ are Hermitian, the above integral can be written as,
\bea 
I_2&=&-\int_0^t d\tau e^{i\mc{L}(t-\tau)}i\frac{\langle   f(t),\mc{L}QA\rangle }{\langle   A, A\rangle }A\nn
&=&-\int_0^t d\tau e^{i\mc{L}(t-\tau)}\frac{\langle   f(t),f(0)\rangle }{\langle   A, A\rangle }A\nn
&=& -\int_0^t d\tau \kappa(t) A(t-\tau).
\eea
This term relates the dissipation in $A$ with the fluctuations in other fast incoherent degrees of freedom. Such a relation is often termed as {\it fluctuation dissipation theorem}\cite{nyquist_28, callen_51}. It leads to the generalized Langevin equation which can be written as,
\bea 
\frac{d}{dt} A(t)=i\Omega A(t)-\int_0^t d\tau \gamma(t) A(t-\tau)+f(t).
\label{eq:langevin}
\eea 
The force-force correlator $\gamma(t)$ and its Fourier transform have in general, complicated time and hence frequency dependence respectively. Thus it carries the information of the memory or the history of the past scattering events and are termed as non-Markovian processes. This is why the kernel describing the effects of the unprojected or incoherent degrees of freedom is termed as {\it memory function}. For a specific case when $\gamma(t)=\gamma_0 \delta(t)$, i.e. when its Fourier transform is constant, the dynamics becomes memory less and are called Markovian process in statistical mechanics literature\cite{vankampen_book}. Let us now compare the above expression which is obtained from the exact microscopic description, with that of the Langevin Equation used from a phenomenological consideration. In the later case the force $f(t)$ is considered as random with a variance assumed phenomenologically.  In principle, $f(t)$ follows deterministic equation of motion and its exact evolution requires solutions of infinite set of equations as we will discuss in the next section. This is computationally impossible as no system is completely isolated and thus the ``total system'' implies the whole universe! To a good approximation, it is justified to consider force of some suitable order as random variables and solving the above equation to get an effective understanding about the system dynamics. Findings from such probabilistic description fits nicely with experimental findings. This is how an effective random  or probabilistic description out of deterministic microscopic principles can emerge in a complex system. However, the origin of randomness in a purely deterministic system is a subtle issue. For more discussions interested readers can consult\cite{zwanzig_book}. 
\section{Continued fraction description} 
\label{sec:CF} To elaborate on the memory function description further we discuss how any dynamical correlation function can be expressed in terms of the static correlations in  a continued fraction form. It was first shown by Mori, in his seminal work\cite{mori_65-2}.
From Eqn. \ref{eq:langevin} we see that, the time dependence of a general dynamical variable say $A(t)$ is dictated by the correlations of its time derivative $\dot{A}(t)$. Same should hold true for $\dot{A}(t)$ and its higher time derivatives also\cite{okada_95}. This leads to a new type of moment expansion as follows. We can  rewrite the generalized Langevin equation
(Eqn. \ref{eq:langevin}) as,
\be
\frac{d}{dt} A(t)=i\Omega_0 A(t)-\int_0^t d\tau \kappa_1(t) A(t-\tau)+A_1(t).
\ee 
Here $A_1$ is termed as the random force and has the same symmetry as $\dot{A}$. If we write its equation of motion, it will also follow a generalized Langevin equation involving higher time derivative of $\dot{A}$. The equation of motion for a $n^{th}$ order ``force'' $A_n$ becomes,
\be
\dot{A}_n(t)=i\Omega_n A_n(t)-\int_0^t d\tau \kappa_{n+1}(t) A_n(t-\tau)+A_{n+1}(t),
\ee
with $n$-th order frequency and the $n+1$-th order memory kernel
\be 
\Omega_n=\frac{\langle   \mc{L}A_n, A_n\rangle }{\langle   A_n,A_n\rangle },\,\, \kappa_{n+1}(t)=\frac{\langle   A_{n+1}(t), A_{n+1}\rangle }{\langle   A_{n}, A_{n}\rangle }
\ee 
respectively. This recurrence formula can be cast in a single  continued fraction form of the correlation function as follows,
\bea 
&& \int_0^\infty dt e^{-izt}\langle   A(t);A\rangle \nn
&=& \frac{\langle   A;A\rangle }{i(z-\Omega_0)+\frac{\Delta_1}{i(z-\Omega_1)+\frac{\Delta_2}{i(z-\Omega_2)+....}}}.
\label{eq:cont}
\eea 
In the above expression $\Delta_n=\kappa_{n}(0)$.  Here the dynamic property of a system is completely described by its static correlations. The above result is in principle exact. But the exact evaluation needs the knowledge of the static correlations upto infinite order\cite{dupis_67}. However depending on the situation one can truncate the continued fraction at some suitable order and get sensible results\cite{pires_88}. A detailed discussions of its use in various systems are beyond the scope of this article. Interested readers may look at the references \cite{sug_91}, \cite{grigolini_83} and \cite{sardella_91} where dynamic correlations in case of simple metal, Hubbard Model and spin $\frac{1}{2}$ XYZ model respectively are cast in the continued fraction form. 
\section{Application to the electronic transport}
\label{sec:ET}
In the previous sections we described  various aspects of the memory function approach. Thus we set up the stage for using it to study the dynamical transport properties of various electronic systems. Here our focus is on the time evolution of the current operator and the correlation of its various components in a generic many body system. In our discussions on electronic systems, first we assume that the momentum is the only nearly conserved quantity in the system\footnote{It means that the related correlation function has a very slow decay, e.g. of the form $e^{-\frac{t}{\tau}}$, with the relaxation time $\tau\rightarrow \infty$.}. Thus there is only one slow mode associated with this conservation law. We study the momentum relaxation of a charged particle under external perturbation. In this case the projector operator is defined solely in terms of the current operator. This assumption holds good if there is no other slow modes associated with any other conservation law or broken symmetry that couples to the charge degrees of freedom\cite{carsten_02, patel_14, lucas_15, subir_15}. However for simplicity we stick to this picture for the time being and will generalize it in the later subsections.

Now we can start with the expression for memory function as defined in Eqn.\ref{eq:memory}. In certain situations, we can evaluate the expression in the spirit of perturbation theory. Memory function can be viewed as ``the self energy'' of the current-current correlation function. It has an added advantage that such ``self energy'' calculation does not require vertex correction. The later is extremely important and problematic when the current-current correlation is expressed through the renormalized single particle propagators \cite{mahan_book}. Now, for further simplicity, we consider the case of an one component current operator $J$ and replace $\vert A \rangle$ by $\vert J \rangle$. To clarify more, in an electronic system $\vert J \rangle\equiv J\vert \Phi_0 \rangle$, where $\vert \Phi_0 \rangle$ is the electronic ground state. The correlation denoted by $\langle   J(t)|J\rangle $ , can be used to represent correlation of the form $\langle   [J(t),J]\rangle $ without changing the form of the Eqn.\ref{eq:corrl}. We choose the later form, as it is used to describe the response function. We focus on the response of an electronic system under an external electric field and the relevant quantity is dynamic conductivity $\sigma(z)$ and is given in terms of a commutator correlation of the current operators. In this case, Eqn.\ref{eq:corrl} can be written as,
\bea 
\sigma(z)=\frac{1}{z -M(z)/\chi}\chi.
\label{eq:ext-drude}
\eea 
It is to be noted that we assume time reversal invariance so that $\langle  J\vert \mc{L}\vert J \rangle=\langle  \dot{J}  \vert J \rangle=0$, i.e. the generalized frequency vanishes. Here we see that using memory function formalism the dynamic susceptibility can be written in an Extended Drude form, frequently used by the experimentalists\cite{basov_11} to explain any non-Drude dynamic conductivity. Here we see that the memory function or the generalized many particle(two particle in this case) self energy defined by Eqn.\ref{eq:memory} is the most important quantity to determine the dynamic conductivity. Within our simplified picture, it takes the following form,
\bea 
M(z)&=&\langle  J \vert \mc{L}Q\frac{1}{z-Q\mc{L}Q} Q\mc{L} J \rangle \nn
&=&\langle  \dot{J} \vert Q\frac{1}{z-Q\mc{L}Q} Q\dot{J} \rangle.
\eea 
To get some qualitative idea, we can opt for a ``high frequency expansion'' of the above expression. We consider an energy scale  $z_0=\langle   \dot{J}|Q\mc{L}Q\dot{J}\rangle/\chi $. As long as $z_0< < |z|$, we can expand inverse operator $\frac{1}{z-Q\mc{L}Q}$ in a series and can rewrite the memory function as follows,
\bea 
M
&=&\frac{1}{z}\langle  \dot{J} \vert Q\left(1+\frac{1}{z}Q\mc{L}Q+\frac{1}{z^2}Q\mc{L}QQ\mc{L}Q+\cdots\right) Q\dot{J} \rangle \nn
&=& \frac{1}{z}\langle  \dot{J} \vert Q\dot{J} \rangle + \frac{1}{z^2}\langle  \dot{J} \vert Q\mc{L}Q\dot{J} \rangle \nn&& 
+\frac{1}{z^3}\langle  \dot{J} \vert Q\mc{L}Q\mc{L}Q\dot{J} \rangle +\cdots
\label{eq:memory-1}
\eea 
Here we use the fact $Q^2=Q$ and the above expansion can be termed as a high frequency expansion. Its validity will depend on how small or large the $z_0/|z|$ is.  Since the time derivative of two different orders are uncorrelated in a system having time reversal symmetry, i.e.  $\langle  J  \vert \dot{J} \rangle, \langle  \dot{J}  \vert \ddot{J} \rangle =0$, as proved in the Appendix \ref{app:jdotj}, 
\bea 
M(z)&=&\frac{1}{z}\langle  \dot{J}  \vert \dot{J} \rangle
+\frac{1}{z^3}\langle  \ddot{J}  \vert \ddot{J} \rangle+\cdots
\label{eq:memory-exp}
\eea 
Clearly this expansion will hold good in high frequencies and will breakdown below certain energy scale set by the incoherent part of the Hamiltonian. Here $J$ is the current operator. Now its time derivative $\dot{J}=[J, H=H_0+H']=[J,H']$ is proportional to the coupling strength $g$ (say) with the dimension of energy of different interactions. Thus the above expansion can be viewed as an expansion in terms of $\frac{g^2}{z^2}$. For very weak interactions one can truncate the above expression at the first term itself and calculate the conductivity. However this perturbation theory is different form the diagrammatic perturbation theory that incorporates interaction effects through single particle self energy and vertex corrections\cite{gotze_72}.  
\subsection{Weak coupling theory} The memory function formalism was first used in a systematic way to calculate the electrical conductivity in case of simple metals with various interactions by G\"otze and W\"olfle \cite{gotze_72}. Similar approach is used by many others in this context\cite{gotze_71, jindal_77, helman_78, capek_85, argyres_89, jung_07, kyrychenko_07}. Their approach can be summarized as follows.
According to the linear response theory, the dynamical conductivity is defined as\cite{kadanoff_63,zubarev_60,mahan_book,arfi_92},
\be 
\sigma(z)=-i \frac{1}{z} \chi(z) + i \frac{\omega_{p}^{2}}{4\pi z}.
\label{cond}
\ee
Here $\omega_{p}^{2}=4\pi N_{e} e^{2} /m$ is the square of plasma frequency where $e$ electronic charge, $m$ electron mass and $N_{e}$ is the electron density, $z$ is the complex frequency and $\chi(z)$ is the current-current correlation function defined as,
\be 
\chi(z) = \langle  \langle   J; J\rangle\rangle_{z} = i \int_{0}^{\infty} e^{izt} \langle   \left[ J(t), J \right] \rangle,
\label{correlator}
\ee
where $J = \sum_{\textbf{k}} e v(\textbf{k})c_{\textbf{k},\sigma}^{\dagger}c_{\textbf{k},\sigma} $ is the current density and $v(\textbf{k})$ is the velocity dispersion. Here $\left[ J(t), J \right]$ denotes the commutator, $\langle   ... \rangle$ denotes the ensemble average at temperature $T$ and $\langle  \langle   ... \rangle\rangle$ denotes the Laplace transform of the ensemble average.

According to the G\"otze and W\"olfle approach\cite{gotze_72}, the memory function is defined as
\be
M(z)=z\frac{\chi(z)}{\chi_{0}-\chi(z)},
\label{cor}
\ee
where $\chi_{0}$ corresponds to the static limit of correlation function (i.e.  $\chi_{0}=N_{e}/m$)\cite{gotze_72}. Using this, the expression for dynamical conductivity in Eqn.(\ref{cond}) can be cast in an Extended Drude form as,
\be 
\sigma(z)=\frac{i}{4\pi} \frac{\omega_{p}^{2}}{z+M(z)}.
\label{conddi}
\ee
In Ref.\cite{gotze_72}, an expansion for $ M(z) = \frac{z\chi(z)}{\chi_{0}} \left(1+\frac{\chi(z)}{\chi_{0}}-\cdots\right)$ is used. Basis of this assumption is the smallness of the contribution from the interaction part as compared to the kinetic energy of free electrons. Using this expansion and on keeping the leading order term, the memory function $M(z)$ can be written as
\be
M(z)=z\frac{\chi(z)}{\chi_{0}} = z\frac{\langle  \langle   J; J\rangle\rangle_{z}}{\chi_{0}}.
\label{memory66}
\ee
To compute memory function, we need $\langle  \langle   J; J\rangle\rangle_{z}$ which by using equation of motion is
\be 
z\langle  \langle   J; J\rangle\rangle_{z} = \langle  [J,J]\rangle + \langle  \langle  [J,H'];J\rangle\rangle_{z}.
\ee
As the first term of r.h.s is zero, hence the above expression is equivalent to second term which can be further calculated by applying equation of motion.
\be 
z\langle  \langle  [J,H'];J\rangle\rangle_{z} = \langle  [[J,H'],J]\rangle - \langle  \langle  [J,H'];[J,H']\rangle\rangle_{z}.\\
\ee
For $z=0$, $\langle  [[J,H'],J]\rangle=\langle  \langle  [J,H'];[J,H']\rangle\rangle_{z=0}$. Thus, the memory function $M(z)$ becomes
\be
M(z)= \frac{\phi(0) - \phi(z)}{z\chi_{0}}.
\label{memory}
\ee
Here $\phi(z)$ (called as correlation function) is defined as
\be
\phi(z)=\left\langle   \left\langle   \left[J,H' \right]; \left[J,H' \right] \right\rangle \right\rangle_{z}.
\label{gen}
\ee

Here the current operator $J_i=\sum v_i(\mb{k})c^\dag_{\mb{k}\sigma}c_{\mb{k}\sigma} $ and its derivative or ``force'' $A=\dot{J}=[J,H]$. Here the total Hamiltonian has two parts, $H_0$ the free part or the kinetic part and an interaction part. While the first part commutes with the current operator, the later part does not. Thus $A=\dot{J}=[J,H_0+H']=[J,H']$  is determined by the interaction part only and is different for different types of interactions.
Within this approach, they calculated the frequency dependent conductivity with various interactions such as electron-phonon, electron-impurity, electron-magnetic impurity, scattering with localized modes etc. To illustrate their work further we will discuss the simplest  case of electron-impurity interaction. 

In this case the interaction part of the Hamiltonian is given as,
\be 
H'=\frac{1}{N}\sum_j\langle  \mb{k}\vert  U \vert \mb{k'}\rangle c^\dag_{\mb{k}\sigma}c_{\mb{k'}\sigma}.
\ee 
Here $U$ denotes the impurity potential. In this case the $j$-th component of the force operator $A$ is given as,
\be 
A_j=\frac{1}{N} \sum\langle  \mb{k}\vert U \vert \mb{k'} \rangle \left[v_j(\mb{k})-v_j(\mb{k'})\right] c^\dag_{\mb{k}\sigma} c_{\mb{k'}\sigma}.
\ee 
Here $v_j(\mb{k})$ is the velocity of the particle with momentum $\mb{k}$ in the $j-$th direction.  The force-force correlation in the case of free electron is estimated as,
\bea
\phi(z)&=&\langle   \langle   A; A\rangle \rangle _z\nn
&=& (2c/3m^2N) \sum_{\mb{k}\mb{k'}}|\langle  \mb{k}\vert U \vert \mb{k'} \rangle |^2(\mb{k}-\mb{k'})^2\nn
&&\times \frac{f(\epsilon_k)-f(\epsilon_{k'})}{z-\epsilon_k+\epsilon_{k'}}.
\eea 
From this expression the imaginary or the absorptive part of the memory function can be estimated as,
\bea
M''(\omega)
&=& \mc{C}\times\frac{1}{N^2} \sum_{\mb{k}\mb{k'}}|\langle  \mb{k}\vert U \vert \mb{k'} \rangle |^2(\mb{k}-\mb{k'})^2\nn
&&\times \left[f(\epsilon_k)-f(\epsilon_{k'})\right]\delta(\omega-\epsilon_k+\epsilon_{k'})/\omega.
\eea 
For momentum independent $U$, i.e point impurity
\be 
M''(\omega)= \mc{C'} (U\rho_F)^2\epsilon_F,
\ee
where $\rho_F$ and $\epsilon_F$ are the density of states at the Fermi surface and the Fermi energy respectively. For $\w<< \epsilon_F$, the imaginary part of the memory function is independent of the frequency and the result is identical to the Drude result. On the other hand, if the impurities are spatially extended,
\be 
M''(\omega)\equiv \frac{1}{\tau}\sim v_F\int\int d\Omega \sigma_{sc}(1-\cos\theta).
\ee 
In the above expression the differential scattering cross section is defined as,  
\be 
\sigma_{sc}(\Omega)=(\pi k_F)^2|\rho_FU(\mb{k_F}-\mb{k'_F})|^2.
\ee
For interactions with non-magnetic impurity we see that the results are identical to the single particle calculations with vertex corrections. This is indeed a benchmark and major success of this formalism.  In other cases there are deviations from
the Drude formula. They argued that these discrepancies are because of spin-flip scattering in a magnetic field, because of resonance
scattering, because of phonon creation at low temperatures, and because of breaking of the
screening cloud attached to charged impurities respectively. However, this version of the memory function approach to calculate the dynamic conductivity is somewhat limited. It is designed for simple metal, i.e. for weakly-interacting electrons with very weak electron-phonon or electron-impurity or other interactions and non expandable to the cases of strong interactions. Lifting these limitations, as required for more exotic systems like strange metal phase in cuprates near optimal doping, and others need substantial improvements. 

\subsection{Strong coupling extension} A large volume of works\cite{ihle_94, plakida_96, plakida_97, jackeli_98, plakida_99, jackeli_99, vladimirov_05, vladimirov_09, plakida_10, vladimirov_11, plakida_11, plakida_11aa, vladimirov_12} on the applications of the memory function in the case of strongly interacting electronic systems is being done by Plakida and his collaborators based on the mathematical formalism develop by Tserkovnikov\cite{tserkovnikov_82}. We see in the G\"otze and W\"olfle formalism, that the dynamical conductivity or the current-current correlation $\langle   J;J\rangle $ can be calculated with the knowledge of $\langle   \dot{J};\dot{J}\rangle $ which is calculated by simple perturbation theory. However it will be seen in this section that, their approach misses some subtle points which may not affect the results in the perturbative limit but can be problematic in the case of strongly correlated systems. Plakida et al. refined the relation between the memory function and the  $\dot{J}$- $\dot{J}$ correlation as follows. First, a relation between two time retarded Green's function and the Kubo-Mori relaxation function can be established as follows. The two time retarded Green's function for two Heisenberg operators $A$ and $B$ are defined as,
\bea 
G^r_{AB}(t-t')&\equiv& \left\langle   \left\langle    A(t)|B(t')\right\rangle \right\rangle \nn
&=&-i\Theta(t-t')\lt\langle   A(t)B(t')-\eta B(t') A(t)\rt\rangle.\nn
\eea 
Here the step function $\Theta(t)=1$ for $t> 0$ and $\Theta(t)=0$ for $t< 0$. The $\langle   ........\rangle $ represents the thermal average i.e. $Tr e^{-\beta H}(..)$ with $\beta$ as the inverse temperature and $\eta=\pm$ for Bosons and Fermions respectively. The above Green's function follows an equation of motion,
\be
i\frac{d}{dt}\left\langle   \left\langle    A(t)|B(t')\right\rangle \right\rangle =\delta(t-t')\lt\langle   \lt[A,B\rt]_\eta\rt\rangle +\left\langle   \left\langle    \dot{A(t)}|B(t')\right\rangle \right\rangle 
\ee
On the other hand Kubo Mori relaxation function is defined as,
\be 
\Phi_{AB}(t-t')\equiv\left(\left( A(t)|B(t')\right)\right)=-i\Theta(t-t')\lt(A(t)|B(t')\rt).
\ee 
The last expression is called the Kubo-Mori scalar product and is defined as\cite{kubo_book},
\be
\lt(A(t)|B\rt)=\int_0^\beta d\lambda \lt\langle   A(t-i\lambda)B\rt\rangle .
\ee 
It can be shown that \cite{tserkovnikov_82},
\bea
\omega\left(\left( A|B\right)\right)_\omega&=&\left( A|B\right)+\left\langle   \left\langle    A|B\right\rangle \right\rangle _\omega\nn
&=&-\left\langle   \left\langle    A|B\right\rangle \right\rangle _0 +\left\langle   \left\langle    A|B\right\rangle \right\rangle _\omega.
\eea 
Now we apply projection operator technique to the Green's function,
\bea 
G_{k,k'}(t-t')=\lt\langle   \lt\langle   A_k(t)|A^\dagger_{k'}(t')\rt\rangle \rt\rangle.
\eea 
It has an equation of motion
\bea
i\frac{d}{dt}\left\langle   \left\langle    A_k(t)|A^\dagger_{k'}(t')\right\rangle \right\rangle &=&\delta(t-t')\lt\langle   \lt[A_k,B_{k'}\rt]_\eta\rt\rangle \nn
&&+\left\langle   \left\langle    \dot{A}_k(t)|A^\dagger_{k'}(t')\right\rangle \right\rangle.
\label{eq:eqnm}
\eea
Here $\dot{A}_k(t)=[A_k(t), H]$. 
Now we extract the linear term in the equation of motion as
\bea 
i\dot{A}_k(t)=[A_k(t), H]=\sum_q E_{k,q} A_q +Z^{ir}_k.
\eea 
The irreducible part  $Z^{ir}_k$ is defined by the orthogonality condition
\bea 
\lt\langle    \lt[ Z^{ir}_k, A^\dag_{k'} \rt]_\eta \rt\rangle =0.
\eea 
This defines the frequency matrix 
\be
E_{kq}=\sum_{k'}\lt\langle    \lt[[A_k(t), H] , A_{k'} \rt]_\eta \rt\rangle I^{-1}_{k'q},\, I_{k'q}=\lt\langle    \lt[A_k , A_{k'} \rt]_\eta \rt\rangle. 
\ee
Upon Fourier transform, Eqn.\ref{eq:eqnm} gives,
\be 
G_{k,k'}(\w)=G^0_{k,k'}(\w)+\sum_{qq'}G^0_{k,k'}(\w)I^{-1}_{qq'}\lt\langle    \lt\langle    Z^{ir}_{q'}| A^\dagger_{k'} \rt\rangle  \rt\rangle. 
\ee
Here the zeroth order Green's function is given as,
\be
G^0_{k,k'}(\w)=\sum_q\frac{I_{qk'}}{\w\delta_{kq}-E_{kq}}.
\ee 
This defines the excitation spectrum in the mean field approximations.
In order to determine the many body part of the Green's function $\lt\langle    \lt\langle    Z^{ir}_{q'}(t)| A^\dag_{k'} (t')\rt\rangle  \rt\rangle $, one needs to differentiate it with respect to $t'$ and after taking Fourier transform one obtains,
\be
G_{k,k'}(\w)=G^0_{k,k'}(\w)+\sum_{qq'}G^0_{k,q}(\w)T_{qq'}(\w)G^0_{q',k'}(\w).
\ee 
The scattering matrix appeared above is defined as
\be 
T_{kk'}(\w)=\sum_{qq'}I^{-1}_{k,q}\lt\langle    \lt\langle    Z^{ir}_{q}|(Z^{ir})^\dag_{q'} \rt\rangle  \rt\rangle I^{-1}_{q'k'}.
\ee 
Now if we define the self energy as,
\be
T_{kk'}(\w)=\Sigma_{kk'}(\w)+\sum_{qq'}\Sigma_{kq}(\w)G^0_{q,q'}(\w)T_{q'k'}(\w).
\ee 
Then the Green's function can be cast in a Dyson form as,
\be
G_{k,k'}(\w)=G^0_{k,k'}(\w)+\sum_{qq'}G^0_{k,q}(\w)\Sigma_{qq'}(\w)G_{q',k'}(\w).
\ee
This tells that the generalized self-energy or the memory function is given by the proper part of the scattering matrix. Thus it can be written as,
\be 
\Sigma_{kk'}(\w)=\sum_{qq'}I^{-1}_{k,q}\lt\langle    \lt\langle    Z^{ir}_{q}|(Z^{ir})^\dag_{q'} \rt\rangle  \rt\rangle ^{proper}I^{-1}_{q'k'}.
\ee 
If we recall the Dyson equations in the electronic Green's function\cite{mahan_book}, we can easily identify that the $T(\w)$ is the generalized many body or multi-particle scattering matrix  while the memory function or the generalized multi-particle self energy  $\Sigma(\w)$ is given by the proper part of $T(\w)$. Diagrammatically it means that the part of the scattering matrix not connected by a single relaxation function. To clarify this statement, a diagrammatic description for the same in case of single particle electronic Green's function or propagator in a coupled electron-Boson system is presented in Fig.\ref{fig:figT}.
\figT \\
Now to calculate the conductivity, we need to focus on the current current correlations.
In this case the relevant response function $\Phi_{JJ}(\w)$ and the memory function or the corresponding multi-particle self energy $ M(\w) $ is related to each other as,
\be 
\Phi_{JJ}(\w)=\lt(\lt(J|J\rt)\rt)_\w=\frac{\chi_0}{\omega+M(\w)}.
\label{eq:phiM}
\ee 
Here $\chi_0=\chi_{JJ}(0)$ and the memory function has both the real and imaginary parts, i.e., $M(\w+\i\delta)=M'(\w)+iM''(\w)$.
Again following the general procedure as discussed earlier, the time derivative of the response function,
\be
\Phi_{JJ}(t-t')=\lt(\lt(J(t)|J(t')\rt)\rt),
\ee
followed by the Fourier transform gives, 
\be 
\Phi_{JJ}(\w)=\Phi^0_{JJ}(\w)+\Phi^0_{JJ}(\w)T_{JJ}(\w)\Phi_{JJ}^0(\w).
\label{eq:phieqn}
\ee 
Here $\Phi^0_{JJ}(\w)=\frac{\chi_0}{\w}$ and in this case the scattering martix is given as,
\be 
T_{JJ}(\w)= \frac{1}{\chi_0}\lt(\lt(\dot{J}|\dot{J}\rt)\rt)_\w\frac{1}{\chi_0}.
\ee 
In order to express the Eqn.\ref{eq:phieqn} in the form of Eqn.\ref{eq:phiM} we need the following relation between the memory function and the scattering matrix.
\be 
T_{JJ}(\w)=-\frac{1}{\chi_0}\left[M(\w)+M(\w)\Phi^0_{JJ}(\w)T_{JJ}(\w) \right].
\ee 
The above equation tells that the memory function for the electrical conductivity is equivalent to the irreducible or the proper part of the force-force correlator, i.e.,
\be
M(\w)= \frac{1}{\chi_0}\lt(\lt(\dot{J}|\dot{J}\rt)\rt)_\w^{proper}\frac{1}{\chi_0}.
\label{eq:m-plakida}
\ee
These authors use this improved definition of the memory function to calculate dynamical quantities in various system such as t-J model in context of  high temperature superconductors and many others in the references cited at the beginning of this subsection. 

As we see, the evaluation of memory function requires evaluation of $\dot{J}=\lt[J,H'\rt]$ and its correlation. Thus any scheme based on the memory function depends on the interaction part of the Hamiltonian and the algebra followed by the operators corresponding to the observables in a system. Also an evaluation of correlator of the form shown in Eqn.\ref{eq:m-plakida} needs some approximations with suitable justifications.

Plakida et al. consider the $t-J$ model\cite{spalek_77, fazekas_99} for the strongly correlated normal phase of the cuprate superconductors to apply their formalism.  The authors use it to calculate both the optical conductivity and the dynamic spin susceptibility in this model. Due to the involved mathematical complexity, the detail description of the application in this model is beyond the scope of this review. However, the findings from this approach are comparable with the other analytical methods and also comparable with the experiments. Interested reader can look at the  references \cite{plakida_99} and \cite{plakida_11}, where Plakida has nicely reviewed the related works.

Similar approach is also used by some other researchers to calculate the spin susceptibility in  strongly correlated electronic systems\cite{maldague_77, kilian_98, prelovsek_01, sega_03, prelovsek_04, sega_06, sega_06a, sega_09}. Since we focus on the electrical conductivity here, we skip those discussions here.

\subsection{Effects of additional slow modes}
Until now we consider systems where the electrical current of the momentum was the only slow mode. However in many systems we need to consider
many other slow modes which couple to the electrical current. For example in a system where there is charge conservation, along with the drift current (related to the momentum of the charge particles), there is another slow mode namely charge diffusion or diffusion current mode\cite{carsten_02, patel_14, lucas_15, subir_15}. The later is also a slow mode and couples to the electric current. In such a system, the projected space along with the electrical current also contains density fluctuation operators. Following Ref.\cite{carsten_02} we define,
\bea 
J_0(\mb{q})=\rho(\mb{q})\,\,\rm{and}\,\, \mb{J}_1(\mb{q})=\mb{J}(\mb{q}).
\eea 
Due to the charge conservation, the density and the longitudinal part of the current operator are related as,
\be 
\mc{L}J_0=-q\mb{J}^L\label{eq:ceqn}.
\ee 
Since the number of slow modes is more than one, in this case, the memory function takes a matrix form as shown in Eqn.\ref{eq:memory}. Here we consider two slow modes, hence the memory matrix has a $2\times 2$ structure and is defined by four correlation functions, namely $R_{00}=\rho-\rho,\, R_{01}=\rho-\mb{J},\, R_{10}=\mb{J}-\rho$ and $ R_{11}=\mb{J}-\mb{J}$. However, the density fluctuations couple only to the longitudinal part of the current and thus in this case  Eqn.\ref{eq:mem0}  can be written in a set of two decoupled equations as follows,
\bea
\sum^1_{l=0}\left[z\delta_{il}-\sum_s \left[\mc{L}_{is}+M^L_{is}\right]\chi^{-1}_{sl}\right]R^L_{ij}(z)&=&\chi_{ij},\nn
\left[z-\left[\mc{L}_{11}+M^T_{11}\right]\chi^{-1}_{11}\right]R^T_{11}(z)&=&\chi^T_{11}.
\label{eq:mem2}
\eea
Here $\chi_{ij}=\lt\langle   J_i|J_j\rt\rangle $ is the static correlation function. Now the continuity equation \ref{eq:ceqn} and the time reversal invariance in the system tell that $\lt\langle   J_0|J_1^L\rt\rangle \sim\lt\langle   J_0|\dot{J}_0\rt\rangle  =0$. Moreover, since $\mc{L}J_0\propto J_1^L$, the first component of the ``four force'' lies within the projected space and thus its unprojected part i.e. $Q\mc{L}J_0$ is identically zero. This simplifies this picture drastically as it leads to
\be 
M_{00}=M^L_{01}=M^L_{10}=0.
\ee 
Thus only $M_{11}=M$ survives and it can be calculated using Eqn.\ref{eq:memory} and assuming $Q\approx 1$\cite{carsten_02, lucas_15}. Once the memory function is determined suitably, the coupled equation can be solved and the effects of charge diffusion can be discussed in varieties of systems as done in references cited in this section.

In this connection we can mention the works of Lucas and Sachdev\cite{lucas_15, subir_15}. In their work they focus on the magneto-transport in strange metals. Their system is a 2D quantum critical metal under an external magnetic field. They consider systems where electronic quasi-particles are absent. Thus they use memory formalism instead of standard perturbation theory to determine the relevant response functions. In their approach they include the effects of other slow modes such as charge diffusions and heat diffusions in their formalism. 
Within this formalism they present some explanation of recently observed  anomalous behavior in the hall angle in the strange metal phase\cite{chein_91}. In summary this approach contains the essential complexity of a typical non-Fermi liquid. It provides a systematic way of including various slow modes within this approach and thus very promising.
\subsection{Comments on few recent works} In this subsection we will qualitatively discuss few of the present day activities based on memory function formalism. We aim to give a flavor of the present day importance of this formalism and keep the discussion very brief. For details, readers are advised to look at the articles cited in appropriate paragraphs.

{\it  Holographic approach:} Since the memory function formalism does not invoke single particle picture to calculate electrical transport. In principle it can be used to calculate transport properties even when quasi-particle picture is not valid. Recently many researchers used both the holographic ADS-CFT principle to study electronic transport in a similar situation, namely 2d metals near quantum critical point\cite{lucas_15}. These models include the coupling between various slow modes and thus produces various non-Fermi liquid transport behavior. To check the consistency, results are often compared with the complimentary memory function calculations and interestingly they have very good agreements.
 
 {\it Non-equilibrium steady state:} Some of the present authors\cite{das_15} used memory function formalism to study electronic transport in non-equilibrium steady state, where the electron and the phonon temperatures are different. 
They consider non-equilibrium relaxation of electrons due to their coupling with phonons in a simple
metal. In their model electrons are living at a higher temperature than that of the phonon bath,
mimicking a non-equilibrium steady state situation. They show that the dc scattering rate at high temperatures and optical scattering rate
at high frequencies, are independent of the temperature difference between the electrons and the
phonons is found in this work. The present formalism forms a basis which can also be extended to
study hot-electron relaxation in more complex situations

{\it Scattering rates in the gapped system:} In an another work two of the present authors\cite{bhalla_15} calculated Generalized Drude scattering (GDS) rate for the case of electron-phonon scattering in metals with a gap in the electronic density of states at the Fermi energy. The resulting GDS is compared with a recent one by Sharapov and Carbotte \cite{sharapov_05} obtained through a different setup. They find good agreement between the two at  finite frequencies. However,
there are  discrepancies in the dc scattering rate which are severe at high temperature which they attribute to some assumptions made in the Sharapov and Carbotte  formalism.
 \subsection{Future directions}
 As mentioned by Lucas and Sachdev in their work \cite{lucas_15, subir_15} a lot of work is needed in incorporating various slow modes in a generic electronic system which often lead to non-Fermi liquid behavior. Depending on the systems under consideration, one needs to include slow modes occurring from various broken symmetries related to  spin density wave, charge density wave, superconductor, nematic transitions in a systematic 
 way. \\ 
Also we notice that the most of the studies within memory function formalism suffers from the lowest order perturbative evaluation of the required two particle correlation function. This is done without much justifications, particularly in case of strong correlations. One needs to improve upon this by considering higher order terms in the the continued fraction representation (Eqn.\ref{eq:cont}) or the high frequency expansion (Eqn.\ref{eq:memory-exp}) as shown in the previous sections. Present authors are involved in such studies.
\section{Discussions}
\label{sec:diss}
 As discussed in this review, though the memory function formalism originated in context of non-equilibrium statistical mechanics, it is being used recently as an important tool to calculate various dynamic transport quantities in various interacting systems. From a theorists point of view, this formalism directly deals with the two particle correlations and thus the existence of electronic quasiparticle is not an essential ingredient here. This is the main advantage of this formalism compared to the single particle perturbation theories which fails in making good predictions in these systems\cite{mahan_book}. Moreover, in this formalism dynamical conductivity can be cast in an  Extended Drude form. The later has a structure as predicted by Drude in case of non-interacting electrons, but with a frequency dependent scattering rate and mass enhancement factor. This form becomes very convenient for experimentalists to estimate the deviation of their data from the simple Drude expression for metals. Thus this method is becoming popular to both the communities. In this article, we summarize the foundation of the memory function formalism. We review its applications in transport studies of various electronic systems in detail. Also we critically examine the approximations used within this formalism in various works and discuss the possible improvements. This review brings all the necessary details of the memory function formalism together at one place. We hope that the present review will be  useful to whoever works in this area, particularly the newcomers.
 \section*{Acknowledgement} We thank Pradeep Kumar for many helpful discussions.
\appendix
\section{Some useful relations}
Here we mention few useful relations. The detailed derivation of them can be found in the reference\cite{tserkovnikov_82}. First we define the Kubo-Mori scalar product as,
\be
\left(A(t), B\right)=\int_0^\beta d\lambda \left\langle   A(t-i\lambda)B\right\rangle .
\ee
Here $\langle   ......\rangle $ represents an equilibrium thermal average. Next, we define Greens function for the above scalar product as follows,
\be 
\left(\left(A(t), B\right)\right)_z=\int_0^\infty dt e^{izt} \left(A(t), B\right).
\ee 
The commutator Green's function is defined as,
\be 
\left\langle   \left\langle   A(t), B\right\rangle \right\rangle _z=\int_0^\infty dt e^{izt}\left\langle   [A(t), B]\right\rangle .
\ee 
These two Green's functions are related as,
\be
z\left(\left(A(t), B\right)\right)_z=\left\langle   \left\langle   A(t), B\right\rangle \right\rangle _z-\left\langle   \left\langle   A(t), B\right\rangle \right\rangle _{z=0}.
\ee
We also have the following relations,
\be
\left(\left(i\dot{A},B\right)\right)_z=\left(\left(A(t), -i\dot{B}\right)\right)_z=\left\langle   \left\langle   A, B\right\rangle \right\rangle _z.
\ee
\be
\left(i\dot{A},B\right)= \left(A(t), -i\dot{B}\right) =\left\langle   [A, B]\right\rangle. 
\ee
Here we see, if $B=A$,
\be
\left(i\dot{A},A\right)= \left(A(t), -i\dot{A}\right) =\left\langle   [A, A]\right\rangle =0.
\ee
\label{app:jdotj}

\end{document}